\documentstyle[aps,prb,epsfig]{revtex}

\begin{document}



\title{Electron transport and intrinsic mobility limits in two-dimensional
electron gases of III-V nitride heterostructures}

\author{Debdeep Jena\thanks{Electronic Mail: djena@engineering.ucsb.edu}
 Yulia Smorchkova,  Chris Elsass,  Arthur C. Gossard, and Umesh K. Mishra}
\address{Department of Electrical and Computer Engineering, University of California,
Santa Barbara, CA 93106}

\maketitle

\begin{abstract}

Electron transport studies for AlGaN/GaN two-dimensional electron
gases is presented. Novel defects in the III-V nitrides are
treated theoretically for two-dimensional transport. Theory of
electron scattering by charged dislocation lines is presented for
realistic two-dimensional electron gases.  The theory lets us draw
new conclusions about the nature of charges residing in the
dislocation in the III-V nitrides. The theory leads to the
recognition of the fact that current state of the art highest
mobility values are limited intrinsically and not due to removable
defects.  Ways of bypassing the intrinsic limits to achieve higher
mobilities are proposed.

\end{abstract}


\section{Introduction}

Over the last decade, the III-V nitrides have matured as a
technologically important semiconductor system.  Various
optoelectronic products based on the nitrides have already gone
commercial \cite{nakamura},\cite{Yifeng}; the material system
holds abundant promise for more. Along with the technological
importance, the nitrides offer a rich diversity of new physical
phenomena such as large polarization fields and large band offsets
\cite{bernardini}.  These novel phenomena present a new way of
creating two-dimensional electron gases (2DEGs) at
heterointerfaces - without intentional modulation doping.  M-Plane
growth on tetragonal LiAlO$_{2}$ has been achieved \cite{patrick},
thus giving the experimentalist control over polarization fields.
Recently, integral as well as fractional quantum hall effects have
been observed in AlGaN/GaN 2DEGs, confirming that the 2DEGs are of
high purity \cite{manfra}.

However the measured electron mobilities are still much lower than
their counterparts in the AlGaAs/GaAs 2DEGs \cite{yulia}.  It is
now well known that low temperature electron mobility in the
purest AlGaAs/GaAs 2DEGs is limited by Coulombic scattering by
remote ionized donors \cite{sakaki}.  However, 2DEGs in AlGaN/GaN
have been created without any intentional modulation doping; still
the low temperature electron mobilities are much lower than the
AlGaAs/GaAs 2DEGs.  We present a detailed analysis of transport in
the AlGaN/GaN 2DEG, and pinpoint the sources of scattering that
are currently limiting low temperature electron mobilities.

\section{2DEG Transport theory}

We first determine the transport regime to choose the correct
theoretical treatment.  Low-temperature transport in the nitride
heterostructure 2DEGs is characterized by long mean free paths;
typical numbers are $L \approx 0.5\mu m$ for a 2DEG of density
$n_{2D}=10^{13}/cm^{2}$ and $\mu=10,000cm^{2}/V.s$.  The
Ioffe-Regel criterion $k_{F}L>>\frac{1}{2\pi}$ is exceeded many
times over, which is characteristic of diffusive rather than
hopping transport.

We point out an interesting fact : the characteristic
dimensionless parameter $r_{s}=E_{e-e}/E_{F}$ (the ratio of
electron-electron potential energy and the kinetic energy), which
is a measure of the importance of many-particle interactions,
ranges from $r_{s}=0.2$ for AlGaN/GaN 2DEG density
$n_{2D}=10^{13}/cm^{2}$ to $r_{s}=2$ for 2DEG density
$n_{2D}=10^{12}/cm^{2}$.  This means that for AlGaN/GaN 2DEGs with
low 2DEG densities, many body effects such as exchange and
interaction can play important effects.  This sensitivity stems
from the relatively heavy electron effective mass
($m^{*}=0.2m_{0}$, $m_{0}$ being the free electron mass) compared
to other III-V semiconductors.  However, electron-electron
scattering is elastic. Drift in response to applied external
fields cannot be impeded by this form of scattering, since energy
lost by one electron in a scattering event is gained by the other
and both electrons contribute to the drift current.  When combined
with other scattering mechanisms, this form of scattering may
affect the calculated mobility; however, we neglect such effects
under the (qualified) assumption \cite{flensberg} that such
corrections are indeed small.

So the treatment is best affected by neglecting localization and
many-particle effects altogether, and treating electron
wavefunctions as extended single particle Bloch states.  Hence,
transport is treated in the traditional regime of the
Born-approximation.  Scattering rate from a state $|k>$ to a state
$|k'>$ is evaluated using Fermi's Golden Rule,

\begin{equation}
S(k,k')=\frac{2\pi}{\hbar}|H_{kk'}|^{2}\delta(E_{k'}-E_{k})
\end{equation}

where $\hbar$ is the reduced Planck's constant,
$H_{kk'}=<k'|V(r)|k> \cdot I_{kk'}$ is the product of the matrix
element $<k'|V(r)|k>$ of the scattering potential $V(r)$ between
plane wave states $<r|k>=1/\sqrt{A}e^{ik_{\perp}r_{\perp}}\chi(z)$
and the matrix element $I_{kk'}$ between lattice-periodic Bloch
functions. Owing to the wide bandgap of the nitrides, the matrix
element $I_{kk'}\approx 1$, the approximation holding good even if
there are large non-parabolicities in the dispersion
\cite{askerov}.

Once the matrix element is determined for different forms of
scatterers, transport scattering time is evaluated by summing over
all the available final states

\begin{equation}
\frac{1}{\tau(k)}=N_{2D} \Sigma_{k'} S(k',k)(1-\frac{{\bf k' \cdot
k}}{|k|^{2}}).
\end{equation}

where $N_{2D}$ is the total number of scatterers in the 2D area
$A$.  Converting the summation to an integral over the
quasi-continuous wavevector states and exploiting the degenerate
nature of the carriers for averaging $\tau(k)$ (all electrons
carrying current have energy $E\approx E_{F}$), the measurable
transport scattering rate $\tau$ reduces to the simple form
\cite{davies}

\begin{equation}
\frac{1}{\tau}=n_{2D}^{imp} \frac{m^{*}}{2\pi \hbar^{3}k_{F}^{3}}
        \int_{0}^{2k_{F}} |V(q)|^{2}
        \frac{q^{2}}{\sqrt{1-(\frac{q}{2k_{F}})^{2}}}
\end{equation}

where $n_{2D}^{imp}=N_{2D}/A$ is the areal density of scatterers,
$m^{*}$ is the conduction band electron effective mass,
$k_{F}=\sqrt{2\pi n_{2D}}$ is the Fermi wavevector, $n_{2D}$ being
the 2DEG density.  Also, $V(q)=V_{0}(q)F_{nm}(q)$, where
$V_{0}(q)=<k'|V(r)|k>/\epsilon_{2D}(q)$ is the screened matrix
element for a perfect 2DEG ($|\chi(z)|^{2}=\delta(z)$), and
$F_{nm}(q)$ is a form factor that takes the finite extent of the
realistic 2DEG along the $z$ direction into account.  For low
temperature transport in the lowest subband (the electric quantum
limit), we denote the form factor by $F_{11}(q)=P_{0}$.

For accurate evaluation of scattering rates, the finite extent of
the 2DEG along the $z$ direction must be accounted for.  The exact
nature of the spatial extent can be evaluated from a
self-consistent solution of Schrodinger and Poisson equations
using the charge control model for the device structure.  Such a
solution is shown in Figure 1 for an Al$_{0.27}$Ga$_{0.73}$N/GaN
2DEG \cite{1DPoisson}.  However, for analytic evaluation of
scattering rates, the Fang-Howard variational wavefunction is a
better candidate, and has been used successfully for transport
calculations in the past. The form of the function is

\begin{eqnarray}
\nonumber
\chi(z)=0, z<0 \\
\chi(z)=\sqrt{\frac{b^{3}}{2}} z e^{-\frac{bz}{2}}, z\geq 0.
\end{eqnarray}

where $b$ is a variational parameter.  The parameter is chosen
such that it minimizes the energy \cite{davies}; this is achieved
when $b=(33 m^{*} e^{2} n_{2D} / 8 \hbar^{2} \epsilon)^{1/3}$,
where the sybols have their usual meaning.  This form of the
wavefunction leads to a form factor

\begin{equation}
P_{0}=\eta^{3}=(\frac{b}{b+q})^{3}
\end{equation}

for transport in the electric quantum limit.  Screening by free
carriers in the 2DEG is also affected due to the finite extent.
This is reflected in another form factor $G(q)$ entering the
long-wavelength ($q\rightarrow 0$) 2D dielectric function
$\epsilon_{2D}(q)=1+\frac{q_{TF}}{q}G(q)$.  Here
$G(q)=\frac{1}{8}(2\eta^{3}+3\eta^{2}+3\eta)$ \cite{afs}, and
$q_{TF}$ is the Thomas-Fermi wavevector.  For a perfect 2DEG,
$\eta \rightarrow 1$, and both form factors reduce to unity.

\section{Defects and Scattering sources}

Transport theory relies on the accurate identification of defects
that lead to scattering of conduction electrons.  Transport in
AlGaAs/GaAs 2DEGs is well understood.  The list of scattering
sources comprises of {\em ionized impurities}, {\em interface
roughness} at the AlGaAs/GaAs barrier, {\em alloy scattering} due
to penetration of the 2DEG wavefunction into the barrier, and {\em
phonons}.  In AlGaN/GaN 2DEGs, all these scattering mechanisms are
present, in addition to some others.  Charged dislocations and
charged surface donors are the new defects, and their effects on
transport for 2DEGs have received little attention.

We find the scattering rates arising out of these new defects as
well as the traditional defects for the AlGaN/GaN 2DEGs. Realistic
values of impurity concentrations specific to AlGaN/GaN samples
got from experimental characterization techniques are incorporated
in the analysis.

\subsection{Dislocations}

Due to the large lattice mismatch of GaN with the substrates on
which it is epitaxially grown (SiC and Sapphire), $N_{disl}=1-100
\times 10^{8}/cm^{2}$ dislocations are typically formed.  These
dislocations originate from the nucleation layer at the interface
of the substrate and GaN, growing up to the surface without
termination.  The dislocations are observed to grow along the
(0001) direction, piercing through interfaces.

The model used for the electrical nature of dislocations to study
it's effect on transport is that of a charged line -  the
dislocation is modeled as a line of dangling bonds, which
introduce states in the energy gap.  The dangling bonds are
separated by a lattice constant $c_{0}$ along the (0001)
direction.  This model was introduced by Read and Shockley
\cite{read}, and has been successful in explaining electron
mobility in bulk semiconductors, including GaN.  The lineal charge
density is given by $\rho_{L} = ef/c_{0}$, where $e$ is the
electron charge, and $f$ is the occupancy number for each
available site.

There has been some controversy regarding the electrical activity
of dislocations, as well as for the value of the occupation
function $f$, given that dislocations are charged.  Through
scanning capacitance microscopy measurements, Hansen {\em et. al.}
\cite{pete} showed that dislocations in GaN are electrically
charged.  Brazel {\em et. al.}\cite{elke} and recently, Hsu {\em
et. al.}\cite{hsu} have shown that dislocations offer highly
preferential localized current paths. Additionally, Kozodoy {\em
et. al.}\cite{peter} showed a direct relationship of reverse
leakage currents in GaN junction diodes to the number of
dislocations.  Schaadt {\em et al}\cite{edyu} confirm the notion
of charged dislocations from their scanning capacitance voltage
measurements, and are also able to predict the amount of charge on
the dislocations and the screening lengths around the charged
lines.

Hence, experimental evidence strongly suggests that dislocations
introduce states in band gap.  However, the controversy has been
in theoretical studies.  Elsner {\em et. al.} calculated from both
ab initio local density functional methods and density functional
tight binding methods \cite{elsner1} for both pure edge and pure
screw type dislocations.  For screw type dislocations, the authors
found deep states in the bandgap.  They found no deep states in
the gap for pure edge type dislocations.  Calculations of Wright
and Furthm\"{u}ller \cite{wright1} and Wright and Grossner
\cite{wright2}, however, show that for both AlN and GaN, edge
dislocations introduce electronic states in the gap.  Leung {\em
et. al.}\cite{leung} did an energetics study of the occupation
probabilities for the states introduced by threading edge
dislocations, drawing upon the theoretical results of Wright {\em
et. al.}.  They find that electronic states introduced in the
bandgap by threading edge dislocations can be multiply occupied,
and the probability of occupation of sites is a function of the
background doping density.

Scattering by charged dislocations in bulk semiconductors has been
treated by several authors \cite{podor}, including for GaN
\cite{look}, \cite{weimann}.  However, scattering by dislocations
for reduced dimensional electron systems had not been analyzed
until recently \cite{myfirstpaper}.  We recently derived the
scattering rates for a perfect two-dimensional electron gas (i.e.,
with zero spatial extent in the z-direction) for charged threading
edge dislocations. We derive scattering rates in this work for a
realistic two-dimensional electron gas with a finite spatial
distribution in the z-direction.  Figure 2 captures the essential
model for the problem.

For simplicity, we assume the dielectric constants in the barrier
and GaN to be same $\epsilon_{b}(AlGaN) \approx \epsilon_{b}(GaN)
= \epsilon$. The dislocation line charge is written as
$\rho_{L}=ef/c_{0}$, where $e$ is the electron charge, $c_{0}$ is
the lattice constant in GaN along the (0001) direction, and $f$ is
the fraction of occupied acceptor states introduced by the
dislocation.  The Fourier transform of the screened potential
experienced by the 2DEG due to a differential charge element
$dQ=\rho_{L}dz$ located a distance $z$ away from the interface is
given by \cite{afs}

\begin{equation}
dU(q)= \frac{e \rho_{L}}{2 \epsilon} \frac{P_{0} e^{-qz} dz}{q+q_{TF}G(q)}.
\end{equation}

$P_{0}$ and $G(q)$ were introduced earlier.  $q_{TF}$ is the
Thomas-Fermi screening wavevector for 2D systems, given by
$q_{TF}=2/a_{B}^{*}$, $a_{B}^{*}$ being the effective Bohr radius
in 2D.  The total potential seen at the 2DEG is got by summing
potentials of all differential elements,

\begin{equation}
V(q)=\frac{e \rho_{L}}{2 \epsilon} \frac{2 P_{0}}{q+q_{TF}G(q)}.
\end{equation}

The scattering rate is evaluated by using Equation [3] as

\begin{equation}
\frac{1}{\tau_{disl}}=N_{dis} \frac{m^{*}}{2 \pi \hbar^{3} k_{F}^{3}}
    \int_{0}^{2k_{F}} dq |V(q)|^{2} \frac{ q^{2} }{ \sqrt{ 1-(\frac{q}{2k_{F}})^{2} } }.
\end{equation}

With the substitution $u=q/2k_{F}$, we reduce the expression to

\begin{equation}
\frac{1}{\tau_{dis}}= \frac{ m^{*} e^{2} \rho_{L}^{2} N_{dis} }{ \hbar^{3} \epsilon^{2} }
            \frac{I(n_{2D})}{4 \pi k_{F}^{4}}.
\end{equation}

The dimensionless integral $I(n_{2D})$ depends on the 2DEG
density. When $b \rightarrow \infty$ with $a=q_{TF}/2k_{F}$, the
integral factor reduces to

\begin{equation}
I(a(n_{2D}))=\frac{\sqrt{1-a^{2}}+a^{2}ln(\frac{1-\sqrt{1-a^{2}}}{a})}{a(1-a^{2})^{\frac{3}{2}}}
\end{equation}

which is the expression for a perfect 2DEG with no spatial extent
in the $z$ direction.

We find the assumption in previous works for bulk GaN scattering
\cite{look},\cite{myfirstpaper} that $f=1$ predicts a lower low
temperature mobility than the highest values reported till date.
The energy calculations by Leung {\em et. al}\cite{leung} show
that typically only $10-50\%$ of the states will be occupied
$(f=0.1-0.5)$ for a background donor density of $N_{d}\approx
10^{16}/cm^{3}$ and dislocation densities in the
$10^{8}-10^{10}/cm^{2}$ range (which is typical of high purity
molecular beam epitaxy samples).  Yu {\em et. al.} \cite{edyu}
reported $f=0.5$ from their scanning capacitance-voltage
measurements.  This makes dislocations much more benign as
scatterers than initially thought - note that $f^{2}$ appears in
the denominator in the expression for mobility. Electron mobility
due to scattering by charged dislocations {\em alone}
$\mu_{disl}=e \tau_{disl}/m^{*}$ is shown in Figure [3] as a
function of 2DEG carrier density.  We show the values calculated
for $f=0.1,0.5,$ and $1$.  The dependence on 2DEG density is found
to be $\mu_{disl} \propto n_{2D}^{1.34}$ (for a perfect 2DEG, the
dependence was found to be $\mu_{disl} \propto
n_{2D}^{\frac{3}{2}}$) \cite{myfirstpaper}.  A useful numerical
formula that can be incorporated in Monte-Carlo techniques for
treating dislocation scattering limited electron mobility in 2DEGs
is

\begin{equation}
\mu_{disl}=58667 \times (\frac{10^{8}}{N_{disl}}) \times (\frac{n_{2D}^{1.34}}{10^{12}})
                \times (\frac{1}{f^{2}})
\end{equation}

where $N_{disl},n_{2D}$ are in $cm^{-2}$, $f$ is the occupation
function for filled states, and the mobility is in $cm^{2}/V \cdot
s$.  There is no temperature dependence, reflecting the degenerate
nature of the carriers.  Also, at a dislocation density of
$10^{9}/cm^{2}$, the mean spacing between the dislocations is
$\approx 300 nm$, whereas the Fermi wavelength for a 2DEG
concentration of $10^{13}/cm^{2}$ is $\lambda_{F} \approx 10 nm$.
This implies that there can be no interference effects in
transport - thus the approximation that the scatterers are
randomly distributed is a good one.

\subsection{Ionized impurity}

The spirit of the HEMT 2DEG is a spatial separation of the 2DEG
from the ionized donors, thus reducing scattering and improving
electron mobility.  State of the art AlGaN/GaN systems have
$N_{back}=10^{16}/cm^{3}$ unintentional residual background
donors. These donors have been detected by secondary ion mass
spectroscopy (SIMS) measurements to be oxygen and silicon atoms
that incorporate during the growth process.  Transport scattering
rate due to a homogeneous background donor density of $N_{back}$
is given by

\begin{equation}
\frac{1}{\tau_{back}}=N_{back} \frac{m^{*}}{2 \pi \hbar^{3} k_{F}^{3}}
            (\frac{e^{2}}{2 \epsilon})^{2}
            \int_{0}^{2k_{F}} dq
                \frac{P_{0}^{2}}{(q+q_{TF}G(q))^{2}}
                \frac{q}{\sqrt{1-(\frac{q}{2k_{F}})^{2}}}.
\end{equation}

The absence of intentional modulation donors implies that there is
a different source of 2DEG electrons. We point out that
polarization by itself {\em cannot} supply electrons, since on the
whole, it is charge neutral.  What polarization {\em can} do
however, is induce the transfer of electrons from any other
possible source to satisfy requirements of electrochemical
equilibrium. Background impurity densities cannot be the source of
the 2DEG electrons since their concentration is too low for
providing the high density 2DEG densities observed.

Electrons in the 2DEG for nitride heterostructures are believed to
be supplied by surface donor states.  The positive surface charge
forms a sheet-charge dipole with the 2DEG which tries to
neutralize the polarization dipole in the AlGaN layer.  Rizzi {\em
et. al.} have confirmed the presence of surface donor states from
XPS measurements \cite{rizzi}.  These experiments confirm the
suspicion that the surface states are charged, and are responsible
for forming the 2DEG, as inferred from a purely charge control
analysis by Ibbetson \cite{ibbo}.

The surface donor states will form a source of scattering
identical to a delta-doped remote donor layer. Thus, the transport
scattering rate for the surface donors can be treated in the same
way as an ordinary delta-doped donor layer.  If the sheet density
of such donors is $N_{s}$, and is at a distance $d$ form the
heterostructure interface (note that this is also the thickness of
the AlGaN barrier), the scattering rate is given by

\begin{equation}
\frac{1}{\tau_{r}}=N_{s} \frac{m^{*}}{2 \pi \hbar^{3} k_{F}^{3}}
            (\frac{e^{2}}{2 \epsilon})^{2}
            \int_{0}^{2k_{F}} dq
                \frac{P_{0}^{2}e^{-2q|d|}}{(q+q_{TF}G(q))^{2}}
                \frac{q^{2}}{\sqrt{1-(\frac{q}{2k_{F}})^{2}}}.
\end{equation}

\subsection{Alloy disorder}

Alloy disorder scattering originates from the randomly varying
alloy potential in the barrier. This form of scattering is known
to be the mobility limiting mechanism for 2DEGs confined in an
alloy channel such as in InGaAs/GaAs heterostructures.  In 2DEGs
confined in binary compounds, alloy scattering occurs as a result
of the finite penetration of the 2DEG wavefunction into the
barrier; the scattering rate is given by \cite{bastard_alloy}

\begin{equation}
\frac{1}{\tau_{alloy}}=\frac{m^{*}\Omega_{0}(\delta V)^{2}x(1-x)}{e^{2}\hbar^{3}}
                        \times \frac{\kappa_{b}P_{b}^{2}}{2}
\end{equation}

where $\Omega_{0}$ is the volume associated with each Al(Ga) atom,
$\delta V$ is the difference in potential by replacing a Ga atom
by Al which is taken to be the conduction band offset between the
AlN and GaN, $\kappa_{b}=2\sqrt{2m^{*}\Delta E_{c}(x)/\hbar^{2}}$
is the wavevector characterizing the penetration of the
wavefunction into the AlGaN barrier ($\Delta E_{c}(x)$ is the
conduction band offset), and $P_{b}$ is the integrated probability
of finding the electron in the barrier. Figure 4 shows the
probability $P_{b}$ as a percentage.

In AlGaAs/GaAs heterostructures, this form of scattering is small,
and often negligible\cite{bastard_alloy}.  However, in AlGaN/GaN
heterostructures, the large electron effective mass, the high 2DEG
density and the large alloy scattering potential all combine to
make this form of scattering strong in spite of the confinement in
the binary semiconductor.  Figure [5] shows the alloy scattering
limited electron mobility for a range of 2DEG densities and alloy
compositions.  The current highest mobilities reported in the
AlGaN/GaN 2DEGs are highlighted, and the path to higher mobilities
is clear from the plot.

\subsection{Interface Roughness}

Scattering at rough interfaces can be severe if the 2DEG density is high, since the 2DEG
tends to shift closer to the interface as the density increases.  Interfaces, however, can
be grown nearly atomically flat by modern epitaxial growth techniques.  The
roughness at heterojunction interfaces has been traditionally modeled by a Gaussian
autocovariance function.  Scattering rate by a rough interface with a root mean square
roughness height $\Delta$ and a correlation length $L$ is given by

\begin{equation}
\frac{1}{\tau_{ir}}=\frac{ \Delta^{2}L^{2}e^{4}m^{*} }{ 2 \epsilon^{2} \hbar^{3} }
            ( \frac{1}{2} n_{2D} )^{2}
            \int_{0}^{1} du
            \frac{ u^{4} e^{-k_{F}^{2}L^{2}u^{2}} }
            { (u+G(u) \frac{q_{TF}}{2k_{F}})^{2} \sqrt{1-u^{2}} }
\end{equation}

where the integral is rendered dimensionless by the substitution $u=q/2k_{F}$.

Figure [6] shows how the distance of the centroid of the 2DEG
distribution from the heterojunction interface for different alloy
concentrations varies with the 2DEG sheet density.  The dependence
was calculated from the self-consistent Fang-Howard variational
wavefunction.  The dependence on the 2DEG density is
characteristically much stronger than on alloy composition.
Interface roughness scattering affects transport even in the
presence of a binary barrier (i.e., absence of alloy scattering).
Recently, Yulia {\em et. al.} reported the successful growth of
AlN/GaN heterostructures, which removes alloy scattering
completely \cite{yulia_aln}.  In such high-density samples, the
mobility is interface roughness scattering limited, and the
calculated low temperature mobilities are well explained by our
theory.

A large distribution of interface roughness sites can form
localized states of the 2DEG instead of the extended states for
high density samples; this limit was analyzed by Zhang and Singh,
who proposed that in such a case, transport will require phonon
assisted hopping \cite{jasprit}.  However, since the highest
reported mobilities are for low density samples ($n_{2D}\approx
10^{12}/cm^{2}$) with no temperature dependence of conductivity
for $T\leq 30K$, we believe that transport in the best samples is
by band conduction.

Interface roughness scattering limited transport mobility has a
characteristic $L^{-6}$ dependence for 2DEGs in quantum wells (of
thickness $L$), which can be observed by transport measurements on
quantum wells of different thicknesses.  An interesting feature of
the III-V nitrides is that due to the unscreened polarization
fields in thin epitaxial layers, there is a large band-bending
inside the well even under no external bias.  Hence the 2DEG
samples one interface much more than the other; this acts as an
in-built mechanism to restrict interface roughness effects on
2DEGs confined in thin unscreened quantum wells.

\subsection{Phonons}

Phonon scattering limits electron mobility at temperatures
$T\geq80K$ for 2DEGs.  Scattering by three types of phonons are
known to affect transport - by deformation potential acoustic
phonons, piezoelectric acoustic phonons, and polar optical
phonons.

Scattering by both types of acoustic phonons is considered elastic
for all practical purposes. Scattering rates of the two types of
phonons have been widely studied; we use the form of scattering
rate derived for the Fang-Howard wavefunction \cite{wladek}.

Polar optical phonon (POP) energy for the wurtzite GaN crystal
lattice is higher than other III-Vs $(\hbar \omega_{pop} = 92
meV)$.  Scattering by polar optical phonons is highly inelastic.
Treatment of scattering of electrons confined to two dimensions by
polar optical phonons has proved to be a difficult problem to
solve, and no satisfactory theory exists.  In spite of the highly
inelastic nature of polar optical phonon scattering, an
approximate momentum relaxation rate in 2DEGs was derived by
Gelmont, Shur, and Stroscio , which accurately explains
experimental data over a wide temperature range \cite{gelmont}.
The result (in CGS units) is given by

\begin{equation}
\frac{1}{\tau_{pop}}=\frac{2 \pi e^{2} \omega_{0} m^{*} N_{B}(T) G(k_{0})}
                            {\kappa^{*} k_{0} \hbar^{2} F(y)}
\end{equation}

Here, $k_{0}=\sqrt{2 m^{*} (\hbar \omega_{pop})/\hbar^{2}}$ is the polar optical phonon
wavevector, $N_{B}$ is the bosonic distribution function
$N_{B}(T)=1/(exp[\hbar \omega_{pop}/k_{B}T]-1)$, and $F(y)$ is given by

\begin{equation}
F(y)=1+\frac{1-e^{-y}}{y},
\end{equation}

$y$ being the dimensionless variable $y=\pi \hbar^{2} n_{2D} /
m^{*} k_{B} T$.  For mobility calculations at high temperatures,
we use this expression.

\subsection{Other scattering mechanisms}

Unintentionally doped (UID) GaN exhibits a n-type nature.
Initially, it was attributed to the formation of nitrogen
vacancies.  However, careful study of the energetics of formation
of such vacancies by Van De Walle and Neugebauer has shown that
formation of nitrogen vacancies is energetically unfavorable
\cite{vandewalle}.  Zhu and Sawaki \cite{zhu} derived the
scattering rates by possible nitrogen vacancies and tried
explaining bulk transport properties from their theoretical
calculations.  In light of recent revelations of little chance of
formation, we choose not to include nitrogen vacancies in our
analysis of electron transport.  However, we point out that
vacancy formation energies may be lowered around dislocations; we
leave this question open for further studies.  In an earlier work,
we presented the theory of scattering arising from the coupling of
polarization dipoles with alloy disorder \cite{mydipolepaper}.
Current highest mobility values are much lower than the mobility
limited by dipole scattering; so we do not include that form of
scattering in this work.

\section{Electron Mobility}

Armed with the scattering rates, we are in a position to relate
the theoretical values to experimentally observed transport
measurements.  We consider low-temperature transport first.

\subsection{Low Temperature}

At low temperatures, the different scattering processes act
independently; Matheissen's rule offers a simple way of combining
the effect of all scatterers.  Figure [7] shows a low temperature
mobility ($\mu=e\tau/m^{*}$) map as a function of the 2DEG sheet
density $n_{2D}$.  The calculation was done with a dislocation
density of $n_{2D}=5 \times 10^{8}/cm^{2},f=0.1$, background
density $n_{back}=10^{16}/cm^{3}$, surface donor density
$n_{s}=n_{2D}$ which is required from charge control, alloy
composition $x=0.09$, and interface roughness parameters $L=$10\AA
, $\Lambda=$2\AA. Every source of scattering has been considered,
and the relative effects are clearly visible in the plot.

Total mobility shows a characteristic maximum; at low sheet
densities $n_{2D} \leq 10^{12}/cm^{2}$, charged impurity
scattering from background donors, surface donors and dislocations
limit electron mobility.  Relative concentration of the particular
form of charged impurity will determine which is the dominant
scatterer.  However, as is evident from the calculation, mobility
at typical AlGaN/GaN sheet densities is limited by short range
scatterers due to alloy disorder and interface roughness.  In the
range of 2DEG densities $n_{2D} \geq 10^{12}/cm^{2}$, alloy
scattering or interface roughness scattering dominate, depending
on the nature of the barrier.  The shaded bands labeled 1 and 2
depict the present-day highest experimental values reported.
Samples whose measured values lie in band 1 have alloy barriers of
low alloy composition, whereas those in band 2 are for 2DEGs at
AlN/GaN heterointerfaces.  The importance of short range
scatterers was also pointed out by Hsu and Walukiewicz in their
recent work \cite{wladek_new}.

The mobility limit for $n_{2D}\geq n_{cr}=10^{12}/cm^{2}$ is
`intrinsic' in the sense that removal of charged defects
(dislocations, background impurities, etc) will not be useful in
improving the mobility.  The critical density $n_{cr}$ can be used
as a guideline for designing high mobility 2DEG structures.  We
predict that highest low temperature mobilities will be achieved
for low density ($n_{2D} \approx n_{cr}$) 2DEGs.  For the same
carrier density, a barrier of AlN will have a larger mobility than
an AlGaN barrier.

If carrier densities are lowered below the critical density
$n_{cr}$, the effects of charged impurities will be visible.
Reduction of carrier density in the AlGaN/GaN 2DEG is not
straightforward, since the 2DEG is {\em not} controlled by
intentional modulation doping.  Gating is one way of achieving low
carrier densities; another way is the growth of GaN (or low
composition AlGaN) cap layers on top of the AlGaN barrier layer.
Introduction of acceptors in the barrier can also be used to
reduce the 2DEG density by compensation, though p-doping in the
nitrides is currently not under good control, and it will open
another source of scattering.

\subsection{High temperature}

At high temperatures ($T>100K$), the approximation that various
scattering processes are independent breaks down due to strong
optical phonon scattering.  However, since the total scattering
rate is dominated by optical phonon scattering, using Matheissen's
rule will not cause significant deviations from a more accurate
calculation \cite{wladek}.

We plot the experimental mobility and carrier concentration values
for a high mobility AlGaN/GaN 2DEG in Figure [8].  The sample
growth specifics was reported elsewhere \cite{yulia}.  The
theoretically calculated values of mobility are also plotted for
the particular 2DEG, showing the effect of all scattering
mechanisms.  It is clear that the mobility (which is one of the
highest reported) is limited by alloy scattering at low
temperatures.

For AlGaN/GaN heterostructures that have a parallel conducting
channel, the 2DEG contribution to transport mobility is masked at
high temperatures by transport in the parallel channel.  In such
cases, it is important to note that the  measured hall mobility
has contributions from both the 2DEG and a parallel channel which
is in general of lower mobility.  The parallel channel is through
low mobility 3D carriers thermally activated from unintentional
donors which freeze out at low temperatures.  So pure 2DEG effects
can be seen only at cryogenic temperatures.  This leads to the
discrepancy in measured and theoretical mobility calculated for
pure 2DEGs.  This is evident in Figure [8] - there is an onset of
parallel conduction through carriers in the parallel 3D layer. The
sharp increase in measured carrier concentration is accompanied by
a drop (shaded region) in measured mobility from the theoretical
values, which are calculated for a pure 2DEG. Room temperature
2DEG mobilities exceeding $\mu_{300K}=2000 cm^{2}/V \cdot s$ have
been reported for samples in which parallel conduction is
negligible, or through a high mobility channel \cite{gaska}.
Removal of the parallel conducting layer will help in reaching the
theoretical mobility limits for 2DEGs imposed by POP scattering.

\section{Conclusions}

In conclusion, the insensitivity of high density AlGaN/GaN 2DEG
transport to various charged impurities in general, and charged
dislocations in particular was shown.  The possible charge
configuration of dislocation states was extracted from
experimental mobility values.  Mobility limiting scattering
mechanisms were determined.  Alloy disorder scattering was shown
to be limiting mobilities for AlGaN/GaN 2DEGs at low temperatures.
Interface roughness scattering is the mobility limiting scattering
mechanism for high density 2DEGs at AlN/GaN interfaces.  Ways of
getting around these apparent `intrinsic' limits to achieve higher
mobilities were outlined.  Effect of parallel conduction at room
temperature was shown to affect measured mobility and cause
deviations from the theoretically predicted values.

\section*{Acknowledgments}
The authors acknowledge helpful discussions with W. Walukiewicz,
I. B. Yaacov, P. Tavernier, M. Manfra, and A. Rizzi.




\begin{figure}
\caption{Self consistent solution of Schrodinger and Poisson
equations yields the z- part of the wavefunction of the lowest
subband, $\chi(z)$.  The electronic density variation, given by
$\rho(z)=e |\chi(z)|^{2}$ is plotted along with the lowest subband
energy.  The difference in the slopes of the conduction band
across the heterointerface is due to the sheet charge of the
polarization dipole in the AlGaN layer.}

\label{fig1}
\end{figure}

\begin{figure}
\caption{Model for calculation of scattering rate due to charged
dislocation for a 2DEG with a finite extent along the z-direction.
The thick line depicts the uniformly charged dislocation. }

\label{fig2}
\end{figure}

\begin{figure}
\caption{2DEG mobility due to scattering by charged dislocation
lines alone.  On the left is shown the dependence of mobility on
the 2DEG sheet density at a fixed dislocation density, on the
right is the dependence on the dislocation density at a fixed 2DEG
sheet density.  Plots for three distinct $f$ values is given to
highlight the sensitivity of scattering to the occupation function
for dislocation introduced states in the gap.}

\label{fig3}
\end{figure}

\begin{figure}
\caption{Integrated probability of the wavefunction penetration in
to the AlGaN barrier as a percentage.  The probability can be as
high as 10\% for typical AlGaN/GaN structures.  Note that the
penetration is reduced strongly with increasing Al composition,
owing to the increase in the barrier height. }

\label{fig4}
\end{figure}

\begin{figure}
\caption{2DEG mobility limited by alloy scattering alone.  Since
alloy scattering is sensitive to 2DEG density $n_{2D}$ and the
alloy composition, the 2D surface shows the required compositions
and densities to reduce the effect of alloy scattering in
AlGaN/GaN 2DEGs.  Note that the flat surface on the top is not a
mobility limit, but a cutoff for the plot.  The figure should
serve as a useful guideline for a achieving higher low temperature
mobilities in the AlGaN/GaN 2DEGs, since the current highest
mobilities are alloy scattering limited.}

\label{fig5}
\end{figure}

\begin{figure}
\caption{Distance of the centroid of the charge distribution from
the heterojunction for three alloy compositions.  The 2DEG moves
closer to the interface as the density increases.  The dependence
on the alloy composition is weak for the same 2DEG density.  For
n$_{2D} \geq$ 10$^{13}$/cm$^{2}$, the 2DEG is very close to the
interface, and interface roughness scattering is very severe. }

\label{fig6}
\end{figure}

\begin{figure}
\caption{Low temperature mobility with contributions from all
defects, impurities, disorder, as well as by acoustic phonons at
$T=1K$.  Two distinct regions are shaded - labeled 1 and 2 with
circles.  Region 1 is where the highest mobilities achieved till
date with AlGaN/GaN samples.  Region 2 is the same, but for
AlN/GaN samples.  Note that the mobility limits are intrinsic, and
the only way to increase the mobility in each case is to reduce
the 2DEG sheet density.}
\label{fig7}
\end{figure}

\begin{figure}
\caption{Temperature dependence of mobility is shown.  The circles
indicate the mobility and 2DEG sheet density measured for a sample
with one of the highest electron mobilities achieved till date.
The lines (solid and dotted) are theoretical calculations.  Note
the deviation of experimental values from theoretical calculations
at $T \geq 40 K$ is accompanied by a sharp increase in measured
charge density, indicating the onset of conduction through the
unintentionally doped bulk, which leads to a drop of measured
mobility.  }
\label{fig8}
\end{figure}


\end{document}